\documentclass[conference]{IEEEtran}
\IEEEoverridecommandlockouts
\usepackage{cite}
\usepackage{amsmath,amssymb,amsfonts}
\usepackage{algorithmic}
\usepackage{graphicx}
\usepackage{textcomp}
\usepackage{subcaption}
\usepackage{hhline}
\usepackage[colorlinks]{hyperref}
\usepackage[normalem]{ulem}

\usepackage[table,xcdraw]{xcolor}
\newcolumntype{L}{>{\centering\arraybackslash}m{1.4cm}}
\newcolumntype{C}{>{\centering\arraybackslash}m{1cm}}
\newcolumntype{K}{>{\centering\arraybackslash}m{1.5cm}}
\newcolumntype{D}{>{\centering\arraybackslash}m{0.95cm}}

\def\BibTeX{{\rm B\kern-.05em{\sc i\kern-.025em b}\kern-.08em
    T\kern-.1667em\lower.7ex\hbox{E}\kern-.125emX}}

\usepackage[capitalize]{cleveref}
\usepackage[printonlyused]{acronym} 
\acrodef {BCE}[BCE]{binary cross entropy}
\acrodef {MAE}[MAE]{mean average error}
\acrodef {RMSE}[RMSE]{root mean square error}
\acrodef {MSE}[MSE]{mean squared error}
\acrodef {NMS}[NMS]{non-maximum suppression}
\acrodef {DNN}[DNNs]{deep neural networks}
\acrodef {TDOA}[TDOA]{time difference of arrival}
\acrodef {DOA}[DOA]{direction of arrival}
\acrodef {SSL}[SSL]{Sound source localization}
\acrodef {STFT}[STFT]{short-time Fourier transform}
\acrodef {tgrid}[TG-rep]{Tight grid}
\acrodef {hmap}[HM-rep]{Heat map}
\acrodef {rgrid}[RG-rep]{Refined grid}
\acrodef {CNN}[CNNs]{Convolutional Neural Network}

\acrodef {PK}[PK]{predicted keypoint}
\acrodef {GK}[GK]{groundtruth keypoint}

\acrodef {FP}[FP]{false positives}
\acrodef {FN}[FN]{false negatives}
\acrodef {TP}[TP]{true positives}

\usepackage{multirow}

\begin{document}

\title{
	Learning Multiple Sound Source 2D Localization
}

\author{
    \IEEEauthorblockN{
        Guillaume Le Moing$^{1,2,\dagger}$,
        Phongtharin Vinayavekhin$^{1}$,
        Tadanobu Inoue$^{1}$\\
        Jayakorn Vongkulbhisal$^{1}$,
        Asim Munawar$^{1}$,
        Ryuki Tachibana$^{1}$,
        and Don Joven Agravante$^{1}$
    }
    \IEEEauthorblockA{
        Corresponding authors: guillaume.le\_moing@mines-paristech.fr, pvmilk@jp.ibm.com\\
        $^{1}$IBM Research, Tokyo, Japan\\
        $^{2}$MINES ParisTech - PSL Research University, Paris, France
    }
}

%
%
\IEEEoverridecommandlockouts
\IEEEpubid{\makebox[\columnwidth]{978-1-7281-1817-8/19/\$31.00~\copyright2019 IEEE \hfill} \hspace{\columnsep}\makebox[\columnwidth]{ }}

\maketitle

\IEEEpubidadjcol

%
%
\begin{abstract}

In this paper, we propose novel deep learning based algorithms for multiple sound source localization.
Specifically, we aim to find the 2D Cartesian coordinates of multiple 
sound sources in an enclosed environment by using multiple microphone 
arrays. To this end, we use an encoding-decoding architecture and propose two 
improvements on it to accomplish the task.
In addition, we also propose two novel localization representations 
which increase the accuracy.
Lastly, new metrics are developed relying on resolution-based multiple source 
association which enables us to evaluate and compare different localization 
approaches.
We tested our method on both synthetic and real world data. The results 
show that our method improves upon the previous baseline approach for this 
problem.

\end{abstract}



%
%
\section{Introduction}
\ac{SSL} is an important topic in audio signal processing. Its output is a 
pose/location representation which is then used for a variety of different 
applications such as monitoring of domestic activities, speech enhancement and 
robotics.
The most common approach of \ac{SSL} is \ac{DOA} 
estimation~\cite{conf:nips-ws:chakrabarty2017,
conf:icra:he2018,
conf:interspeech:he2018}.
In \ac{DOA}, the location is represented as one or more angles of the 
sound direction. 
Since the full pose is not represented, it is ambiguous. This ambiguity needs to be 
resolved before it can be used in several applications. For example, in the 
monitoring of domestic activity, a good location representation is the 2D point in a 
floor plan.
The \ac{DOA} output is ambiguous as it points toward a specific angle without 
depth information.
To handle these applications, it is 
possible to combine the result of multiple DOA setups to produce the 
necessary 2D position information.
A detailed review of such methods is given in~\cite{jour:wcmc:cobos2017}. 
However, not all of these methods are suitable in the case of multiple sound 
sources.
A well-known problem of these methods is the so-called \textit{association 
ambiguity}~\cite{jour:signal:ma2006,jour:tasl:Alexandridis2018}, whereby these methods 
produce more
candidates than there are actual sources.
Although some solutions have been 
proposed~\cite{jour:sp:griffin2015,jour:aslp:sundar2018}, this is still an active 
research area.
Differently from \ac{DOA} combination methods, this paper is concerned with 
\ac{SSL} that directly produces a suitable 2D position representation as an output.
Since this is difficult to do directly, in this work we propose and evaluate the 
use of deep learning methods.

Recently, applying \ac{DNN} to acoustic applications has gained 
popularity~\cite{
conf:nips-ws:chakrabarty2017,
tech:dcase:Inoue2018,
conf:slt:takeda2016,
conf:icra:he2018,
jour:sensors:diaz2018,
conf:interspeech:he2018,
conf:mlsp:vesperini2016,
jour:stsp:adavanne2019,
conf:eusipco:adavanne2018}.
A data-driven approach allows to directly learn the complex mapping
from sound features to source locations. 
There are works showing that DNN-based approaches are better 
than classical methods in the presence of noise
and reverberations~\cite{conf:nips-ws:chakrabarty2017}.
However, research on 2D deep acoustic localization is still at an
early stage and there are still several open issues.
An important related work has shown that it is possible to 
output the Cartesian point coordinates directly from raw 
sound~\cite{jour:sensors:diaz2018}, or from preprocessed 
features~\cite{conf:mlsp:vesperini2016}. 
However, because the output structure of a DNN needs to be pre-defined, this 
approach is limited to a fixed number of sound sources which should also be 
reflected in the training data. 
The networks in~\cite{jour:sensors:diaz2018,conf:mlsp:vesperini2016} 
were only designed for a 
single source hence the approach can not handle multiple sound sources.
In~\cite{jour:stsp:adavanne2019}, multiple sound sources are handled by
outputting the coordinates for a predefined number of different sound classes. 
This tackles multiple source localization because different sound classes can 
be active 
simultaneously. However, this approach does not work with simultaneous same-class 
sound sources. 
Another method to handle multiple sources configures the network output as a
finite set of candidate locations~\cite{conf:slt:takeda2016}. 
The output is then trained as a binary classifier whether a source is active or not 
in that particular location.
In~\cite{conf:eusipco:adavanne2018}, a 3D \ac{DOA} is proposed where 
they introduce an intermediate output layer, using a \textit{regression-based} 
approach. However, it still needs a binary classifier to identify active 
sources.
For \ac{SSL}, these \textit{classification-based} methods appear to be the most 
common approach for multiple 
source 2D acoustic localization by deep learning.
A clear limitation of this method is in scaling towards finer resolutions as a 
higher number of output candidates are required.
This paper aims to solve these issues.

%
%
%
%
%
%
%

\begin{figure*}[ht]
    \centering
    \includegraphics[width=0.98\linewidth]{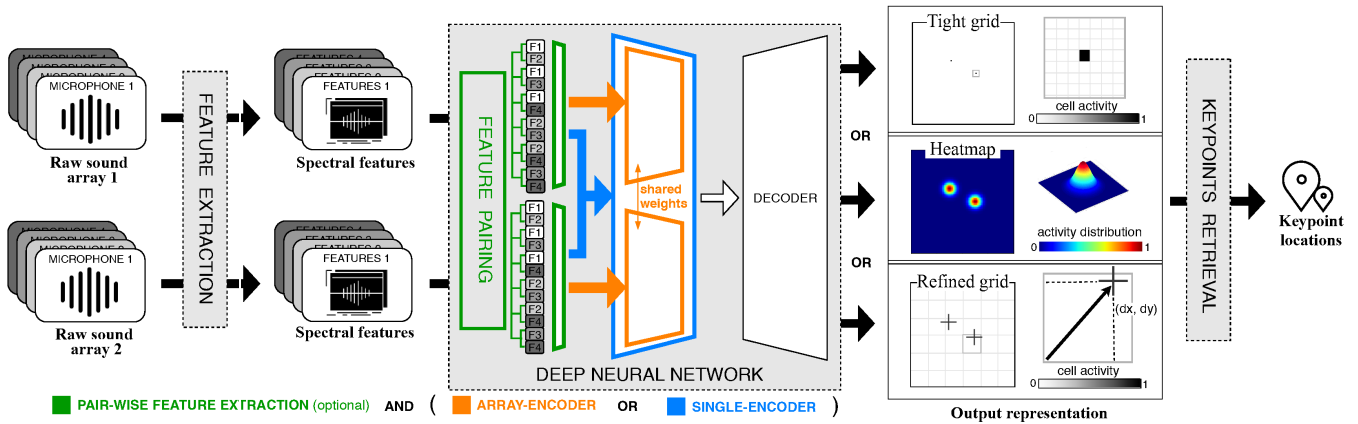}
    \caption{Outline of the localization framework}
    \label{fig:framework}
\end{figure*}

%
%

The main contributions of this paper are:
(1) We propose two novel representations of source location for DNN 
outputs [\cref{ssec:output-representation}].
The first represents locations as probability distributions (heatmap), while 
the second represents locations by combining classification-based grid and 
relative position (refined grid).
The results show that both representations increase the accuracy.
(2) We propose to use an encoding-decoding network architecture for 2D \ac{SSL} 
and provide two improvements: sharing encoders between microphone arrays and 
pairing features among microphones [\cref{ssec:dnn-architecture}].
These improvements allow the network to generalize better when there is 
less training data available.
(3) When there are multiple sound sources, it is difficult to evaluate the 
performance of \ac{SSL} methods because the association between predicted and 
ground truth sources can be ambiguous.
Inspired by~\cite{conf:eusipco:adavanne2018}, we introduce a novel and 
comprehensive evaluation metric 
using resolution-based keypoint association [\cref{ssec:evaluation-metrics}].
(4) Finally, we perform extensive experiments on both synthetic and real world 
data [\cref{ssec:results}].
We also propose a strategy to improve the training of classification-based 
methods with a finer resolution grid for a fair comparison 
[\cref{ssec:output-representation}].
The results show that the proposed methods outperform the existing methods in 
all aspects.

%
%
\section{Multiple Sound Source 2D Localization}

The framework for 2D multiple \ac{SSL} is shown 
in~\cref{fig:framework}. The input is raw sound from multiple microphone arrays 
and the final output is source keypoint locations.
The first step is to extract \ac{STFT} features from all the sound channels.
Similar to~\cite{conf:interspeech:he2018}, we are using the real and 
imaginary 
part of 
the spectrograms, discarding the zero-frequency component. The \ac{STFT} 
features are fed to a neural network which outputs a representation for source locations.
The neural network is trained by supervised learning from labeled data.
Finally, a post-processing step transforms location 
representations into 2D keypoint locations. The next subsections detail the 
major points of our method.

\subsection{Deep Neural Network Architecture}
\label{ssec:dnn-architecture}

The architecture has two main components.
The first component \textit{encodes} spectral features into 
deep features, while the second component \textit{decodes} them to produce the 
target location representations.
For this \textit{encoder-decoder} configuration, the simplest approach would be 
to pass the features of all microphones of all arrays at once through the same 
encoder and then generate location representations by processing the features 
with a decoder.
Here, we propose two improvements for the architecture which are highlighted 
in~\cref{fig:framework}.

\textbf{Array-encoder} Instead of feeding all of the features to the same 
encoder, we 
propose to have a different encoder for each microphone array.
This is similar to classical methods which tackle 2D localization by combining 
features computed at the array level.
The intuition is that the use of multiple encoders helps the model to decompose 
the localization task into simpler subtasks.
Moreover, when two or more microphone arrays have similar properties and are
placed in locations with similar reverberation signatures, the weights of their
encoders can be shared because the deep features learned should be similar and
independent of the array. This allows the encoders to share their training
data, resulting in better generalization with a smaller amount of total
training data.


\textbf{Microphone pair-wise features} Most classical methods process 
microphone signals by pairs to get localization cues.
For instance, some methods localize sources by identifying the delay or the 
difference of attenuation of sound level between pairs of microphones.
We reflect this idea in our architecture by treating the input as pairs of 
microphones rather than a sequence of individual channels.
Specifically, the spectral features are duplicated and rearranged so that all 
combinations of channels in the array can figure side by side.
Then, an additional convolutional layer, with kernel and stride of $2$, is 
added before the encoder to extract pair-wise features.
This modification can be applied either on the original architecture or on top 
of the array-encoder modification.

\subsection{Output Representation}
\label{ssec:output-representation}

The DNN output representations are also illustrated in~\cref{fig:framework} and 
detailed hereafter. We start by explaining a baseline representation, then 
proceed to the two proposed representations that aim to increase the 
localization accuracy.

\textbf{\textbf{}\ac{tgrid}} This is the baseline representation.
The 2D environment is divided into an 
$M$x$N$
grid.
Localization is done at the level of a cell, $c$, by a binary labeling of its 
activity, $A_c$. Active cells, having a sound source inside, are labeled as
($A_c = 1$) 
so that  
($A_c = 0$) 
for inactive cells.
The output is represented by an $M$x$N$x$1$ tensor.
The loss function is a standard \ac{BCE} loss done cell-wise:
\begin{equation}
    \text{BCE}(A_c^{gt}, A_c^{p}) 
    = 
    -A_c^{gt}\log(A_c^{p})
    -
    (1-A_c^{gt})\log(1-A_c^{p})
    ,
    \label{eq:bce}
\end{equation}
where $A_c^{gt}$ is the ground truth label and $A_c^{p}$ is predicted by the 
model.
Applying (\ref{eq:bce}) na\"ively has the problem that precision is directly 
proportional to the output tensor size. That is, a finer grid with more cells 
is required for a higher precision.
This leads to a high class imbalance between active and inactive cells. Class 
imbalance is a common problem in machine learning for 
classification.
To mitigate this, we introduce weights in the loss function to encourage 
predictions:
\begin{equation}
    L_{TG} 
    = 
    \sum_{\text{active }c}\text{BCE}(A_c^{gt}, A_c^{p})
    +
    \lambda \sum_{\text{inactive }c}\text{BCE}(A_c^{gt}, A_c^{p})
    .
    \label{eq:tg-loss}
\end{equation}

\textbf{\ac{hmap}} This representation adopts the same grid resolution as 
that of the \ac{tgrid}. But instead of binary labels, locations are 
represented as probability distributions.
Since the output is distributed across a region of cells, 
the model can gradually learn to predict the target output.
Moreover, \textit{sub-cell} accuracy can be obtained since the center of the 
probability distribution can be anywhere.
In our implementation, the location representation is an aggregation of 2D 
discretized Gaussian distributions centered at the source locations where the 
aggregation function is the maximum function.
A similar representation was adopted in~\cite{conf:icra:he2018} for 1D 
\ac{DOA} estimation.
In order to facilitate the training, the representation is normalized to 
$[0, 1]$. The output is represented by an 
$M$x$N$x$1$ 
tensor, while the loss 
function between predicted and ground truth heatmap is the \ac{MSE}.
At the cell level, the \ac{MSE} between ground truth cell activity $A_c^{gt}$ 
and 
predicted cell activity $A_c^{p}$ is:
\begin{equation}
    \text{MSE}(A_c^{gt}, A_c^{p}) 
    = 
    \frac{(A_c^{gt}-A_c^{p})^2}{\text{number of cells}}
    .
    \label{eq:mse}
\end{equation}
Therefore, the loss for \ac{hmap} is:
\begin{equation}
    L_{HM} 
    = 
    \sum_{c}\text{MSE}(A_c^{gt}, A_c^{p})
    .
    \label{eq:hm-loss}
\end{equation}

\textbf{\ac{rgrid}}
This novel representation combines the \ac{tgrid} approach with direct location 
regression methods~\cite{jour:sensors:diaz2018,conf:mlsp:vesperini2016}.
It is difficult to train \ac{tgrid} for finer grids, while direct regression 
cannot be used with a varying number of sound sources.
Combining them allows the network to overcome both limitations.
This was inspired by the representation in visual object 
detection~\cite{conf:cvpr:redmon2016}.
Similar to \ac{tgrid}, the network first predicts the source location at the 
cell level.
Then, the prediction is refined by regressing the relative position 
$P_c=(dx, dy)$ of the source along x and y axes within the cell.
This 2-step approach allows for a coarser grid.
The output is represented by 
an
$M$x$N$x$3$ 
tensor where the last dimension combines the probability that a cell is 
active and its relative position.
To learn whether a cell is active, we use \ac{BCE} together with weight 
$\lambda_{1}$ to handle the imbalance issue similarly to (\ref{eq:tg-loss}).
For the relative position loss, we compute the \ac{MSE} between predicted 
$P_c^{p}$ 
and ground truth $P_c^{gt}$ relative positions.
Similar to~\cite{conf:cvpr:redmon2016}, the relative position loss is only 
applied to cells containing an active source. We also introduce a weight for the 
relative position loss to balance both classification and relative position objectives.
Combining these we have:
\begin{align} 
    L_{RG} 
    = 
    &\sum_{\text{active }c}\text{BCE}(A_c^{gt}, A_c^{p}) 
    + 
    \lambda_1 \sum_{\text{inactive }c}\text{BCE}(A_c^{gt}, A_c^{p}) 
    \nonumber 
    \\
    &
    + 
    \lambda_2\sum_{\text{active }c}\text{MSE}(P_c^{gt}, P_c^{p})
    .
    \label{eq:rg-loss}
\end{align}

%
\subsection{Keypoint Retrieval}
\label{ssec:post-processing}

During inference the network output is converted into sound source 2D coordinates,
later referred to as keypoint locations.
The method for each representation is described below.

\textbf{\textbf{}\ac{tgrid}} We threshold each cell to identify whether there 
is a sound source inside.
Depending on sound activity, the weighted classification loss, 
(\ref{eq:tg-loss})
can encourage the network to predict the neighboring cells to be active.
Simply thresholding the network output can be insufficient.
To improve this, Gaussian smoothing and \ac{NMS} can be applied.


\textbf{\ac{hmap}} The resulting heatmap is convolved with a Gaussian kernel, 
with the 
same standard deviation as the 2D discretized Gaussian used to produce the 
ground truth. This also filters out high frequency noises.
Then, to recover local maxima, \ac{NMS} is applied such that only the maxima 
with source probabilities over the threshold are selected as keypoint locations.
Finally, a \textit{sub-cell} location is calculated for each keypoint by 
considering the surrounding cell locations of size $10 \times 10$ and then
weight-averaging their probabilities.

%

\textbf{\ac{rgrid}} When the source location is near the edge of a cell, 
the network tends to predict the surrounding cells to be active.
\ac{NMS} is applied to filter these out. As before, only cells with 
a probability higher than a certain threshold are selected as sources.
Taking the predicted relative position of these maxima provides the final 
keypoints.




%
%
\section{Experimental Setup}

\subsection{Dataset}

\begin{figure}[tb]
    \centering
    \includegraphics[width=0.95\columnwidth]{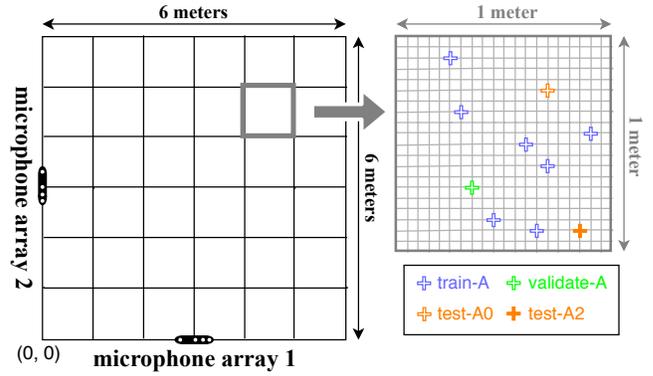}
    \caption{Environment Layout Configuration}
    \label{fig:expsetup}
\end{figure}

We consider as our environment a confined space of $6 \times 6$ meters within 
a large empty office room with a thin carpet floor.
Two linear microphone arrays, from a 
\textit{Kinect Xbox One}\textsuperscript{\textregistered}\footnote{Microsoft; 
http://www.xbox.com/en-US/xbox-one/accessories/kinect}, with a layout 
depicted in~\cref{fig:expsetup}, are used for
recordings.
The sound sources are from a large set of classical, funk and jazz songs. 
From these, $6$s excerpts are taken.
The classical and funk excerpts are played when recording training, 
validation and testing data splits.
The jazz excerpts are used to capture additional testing data to verify the 
generalizability of the model.
One or more sources at various locations can be active at the same time.
When multiple sources are active, we set the minimum distance between sources to be 2 meters.
This limitation eases the evaluation process from the need to separate source pairs that are close to each other.

In this paper, we perform experiments in both a simulated and a real world 
environment with the same layout configuration.
Pyroomacoustics~\cite{conf:icassp:scheibler2018} is used as the simulated environment.
No noise or echo is simulated.
We synthesize the dataset by generating $160$ms audio samples on a $5$x$5$ cm 
resolution grid.
The sound power of the samples vary widely. To prevent including silent 
pauses in the dataset, we added thresholding on the average L1-norm of the 
amplitude.

Capturing and labeling real world audio samples that cover the whole target 
environment with various 
source location combinations requires a lot of effort.
To ease this process, we take advantage of the fact that sound can be 
superposed.
That is, we only record a single source at a time and later combine 
audio samples to cover the multiple source cases.
To get the real world dataset, we begin by dividing a target environment into 
36 cells of $1$x$1$ meter grid. 
Within each cell, we randomly choose a set of sound source positions in a grid 
of $5$x$5$ cm resolution, we then play musical excerpts of a specific genre in 
a random 
order through the speaker of a mobile phone and capture $10$s audio clips.
In each cell, we record $7, 1, 1$ positions for training, validation and testing data splits, respectively.
Audio samples of $160$ms are generated by randomly choosing or mixing one or 
more segments of these audio clips.
We perform this for all data splits independently. For instance, 
the training split has $36\times7=252$ pieces of $10$s audio clips. We randomly 
choose 
and mix audio segments to create $10^5$ audio samples. We refer to this as a 
real world dataset with augmentation.

To confirm that our augmentation for multi-source is valid, we also capture 
samples of the actual case - where multiple sources are present simultaneously 
in the scene. 
It takes a lot of effort to capture and label source locations for each 
data so there are fewer data samples for this case. 
In particular, we only capture this type of data in a limited number of 
location sets (labels) but since the sound excerpt (input) is varying, each 
location provides $10$ \textit{different} samples.
The process for recording this dataset is similar but with multiple sources 
playing at the same time.
%
Finally, we also selected generated segments with sufficient energy similar to synthetic set.
The details for the 
datasets 
are shown 
in~\cref{tab:dataset}.
Datasets with one or two sources contain equal amount of samples for 
both.

\begin{table}[bt]

\begin{center}
\caption{Datasets Used in Experiments}
\label{tab:dataset}  
\begin{tabular}{|l|l|c|c|c|}
\hline
\multicolumn{1}{|c|}{\textbf{Dataset}}                                           
      & \multicolumn{1}{c|}{\textbf{Split}} & \textbf{Excerpts} & \textbf{\# of 
Srcs} & \textbf{Samples}  \\ 
\hline
\multirow{5}{*}{Synthetic}                                                             & train-S                             & classical-funk        & 1 or 2              & 100000            \\ 
\cline{2-5}
                                                                                       & validate-S                          & classical-funk        & 1 or 2              & 5000              \\ 
\cline{2-5}
                                                                                       & test-S0                             & classical-funk        & 1 or 2              & 5000              \\ 
\cline{2-5}
                                                                                       & test-S1                             & classical-funk        & 3                   & 2500              \\ 
\cline{2-5}
                                                                                       & test-S2                             & jazz                   & 1 or 2              & 5000              \\ 
\hline
\multirow{5}{*}{\begin{tabular}[c]{@{}l@{}}Real world with\\Augmentation\end{tabular}} & train-A                             & classical-funk        & 1 or 2              & 100000            \\ 
\cline{2-5}
                                                                                       & validate-A                          & classical-funk        & 1 or 2              & 5000              \\ 
\cline{2-5}
                                                                                       & test-A0                             & classical-funk        & 1 or 2              & 5000              \\ 
\cline{2-5}
                                                                                       & test-A1                             & classical-funk        & 3                   & 2500              \\ 
\cline{2-5}
                                                                                       & test-A2                             & jazz                   & 1 or 2              & 5000              \\ 
\hline
\multirow{3}{*}{Real world}                                                            & test-R0                             & classical-funk        & 1 or 2              & 600               \\ 
\cline{2-5}
                                                                                       & test-R1                             & classical-funk        & 3                   & 300               \\ 
\cline{2-5}
                                                                                       & test-R2                             & jazz                   & 1 or 2              & 600               \\
\hline
\end{tabular}
\begin{scriptsize}
\newline
*youtu.be/ : classical [8f-CFlyNwLE], funk [thUQr7Q1vCY], jazz [oGkkusOMFj0]
\end{scriptsize}
\end{center}
\vspace{-1.0em}
\end{table}


\subsection{Evaluation metrics}
\label{ssec:evaluation-metrics}

Using the corresponding keypoint retrieval method described 
in~\cref{ssec:post-processing}, we obtain a list of 2D coordinates as source 
locations.
To evaluate this result, the two common distance metrics in the \ac{SSL} 
literature are \ac{RMSE}~\cite{jour:tasl:Alexandridis2018} and 
\ac{MAE}~\cite{conf:icra:he2018}.
We choose \ac{RMSE} so as to penalize when the model makes large prediction 
errors. \ac{RMSE} is well suited for single \ac{SSL} as there is a direct 
mapping from a \ac{PK} to a \ac{GK}.
However, in multiple \ac{SSL} where the number of source keypoints is unknown 
beforehand, such a mapping is not straightforward.
Previously, Alexandridis~\textit{et al.}~\cite{conf:mmsp:Alexandridis2018} 
circumvent the issue by allowing one \ac{PK} to be associated with more than one 
\ac{GK}.
However, this method only takes into account the closest \ac{PK} to every 
\ac{GK} and does not reflect mismatch in the number of \ac{PK}s and 
\ac{GK}s.

Instead, we propose to use a resolution-based association method and then use 
precision, recall, and F1 scores as metrics to account for the varying number of 
keypoints.
To compute these, we first match a \ac{PK} to a \ac{GK} if their distance is 
below the resolution threshold.
We then define \ac{TP} as the number of \ac{GK}s that are matched 
(note that we count the number of \ac{GK}s, not the matches); \ac{FP} as number 
of \ac{PK}s minus \ac{TP}; and \ac{FN} as the number of unmatched \ac{GK}s.
Thus, these metrics penalize all possible mismatches between \ac{PK}s and \ac{GK}s.
We consider $2$ resolution values of $0.3$ and $1$ meters to account for fine and coarse localization.
In addition, since precision, recall, and F1 do not account for distance error, 
we also include \ac{RMSE} of the 
\ac{TP}
as another 
metric.
While we propose to use 4 metrics, we relied for the most part on F1-score and \ac{RMSE} to discriminate between methods because they give a global understanding of the localization performance whereas precision and recall provide insights to support the interpretation.

\subsection{Experimental Protocols}
%
%
During experiments, samples consist of two clips of four channel $160$ms audio 
sampled at 16kHz. This is captured with two identical microphone arrays.
For each array we extract \ac{STFT} features 
with Hamming windows of 32 ms with 
$50\%$ overlap, resulting in a shape of $8$x$9$x$256$ 
(channel $\times$ temporal $\times$ frequency), where the real and imaginary 
part of the \ac{STFT} are stacked together in the first axis [$R_1, ..., R_4, 
I_1, ..., I_4$].
The full details are tabulated 
in~\cref{tab:architecture}.
\begin{table}[bt]
\footnotesize\addtolength{\tabcolsep}{-5pt}
\begin{center}
\caption{Deep neural network detailed architecture}
\label{tab:architecture}
\begin{tabular}{|L|CKKCK|}
\hline
Block                                          & Filters             & Kernel                 & Conv type     & Norm     & Activation     \\ \hline
Input                                          & \multicolumn{5}{c|}{Spectral features (one array): \textbf{8x9x256}}      \\ \hline
                                               & \multicolumn{5}{c|}{Pairs of microphones (one array): \textbf{24x9x256}} \\ \cline{2-6} 
                                               & \multicolumn{5}{c|}{Reshape: \textbf{24x9x256} $\rightarrow$ \textbf{9x1x24x256}}  \\ \cline{2-6} 
                                               & 8                   & 2x7                    & conv2d        & bn2d     & LeakyReLU      \\ \cline{2-6} 
\multirow{-4}{1.45cm}{\centering Pair-wise feature extraction} & \multicolumn{5}{c|}{Reshape: \textbf{9x8x12x256} $\rightarrow$ \textbf{96x9x256}}  \\ \hline
& \multicolumn{5}{c|}{\begin{tabular}{|DKKCK|}
  \firsthline
  128 & 1x5 & conv2d & bn2d & LeakyReLU\\
  \hline
  \end{tabular} $* 5$ } \\                                              
                                               & 64                  & 1x3                    & conv2d        &          & LeakyReLU      \\
                                               & 32                  & 1x3                    & conv2d        &          & LeakyReLU      \\
                                               & 16                  & 9x4                    & conv2d        &          & LeakyReLU      \\ \cline{2-6} 
\multirow{-5}{*}{Encoder}                      & \multicolumn{5}{c|}{Reshape: \textbf{16x1x32} $\rightarrow$ \textbf{512x1x1}}      \\ \hline
                                               & 256                 & 3x3                    & dconv2d       & bn2d     & ReLU           \\
                                               & 128                 & {\color{red} 3x3} / {\color{blue} 2x2}      & dconv2d       & bn2d     & ReLU           \\
                                               & 64                  & 3x3                    & dconv2d       & bn2d     & ReLU           \\
                                               & 32                  & 3x3                    & dconv2d       & bn2d     & ReLU           \\
                                               & 16                  & 3x3                    & conv2d        &          & ReLU           \\
                                               & 8                   & 3x3                    & conv2d        &          & ReLU           \\
\multirow{-7}{*}{Decoder}                      & {\color{red} 1} / {\color{blue} 3}       & 3x3                    & conv2d        &          & ReLU           \\ \hline
Output                                         & \multicolumn{5}{c|}{{\color{red} \ac{tgrid} \& \ac{hmap}: \textbf{1x81x81}}\ \ \ \ {\color{blue} \ac{rgrid}: \textbf{3x6x6}}}                         \\ \hline
\end{tabular}
\end{center}
\normalsize
\vspace{-1.0em}
\end{table} 

Experiments are conducted on synthetic datasets and real world datasets separately.
The models trained with synthetic data are used on the synthetic test data (test-S).
Similarly, the models trained with real world augmented dataset are used to predict on both augmented (test-A) and real world (test-R) datasets.
The optimal threshold value for all keypoint retrieval methods are determined 
empirically to be $0.6$, except the improved version of \ac{tgrid} which is set 
to $0.1$ due to a smoothing effect from the Gaussian kernel.
All models were trained for $200$ epochs and the weights that 
result in the best F1-score on the validation data are used.
The ADAM optimizer is used with an initial $\textrm{lr}=0.0001, \beta=(0.5, 
0.999)$ and a batch size of $128$ samples.

%
%
%
%
%
First, we compare output representations by fixing the network architecture.
Our experiments have a square environment so $N=M$ for all representations.
\textbf{\ac{tgrid}} is trained using $\lambda=0.01$ in~\cref{eq:tg-loss}.
To get the keypoint locations, we apply both na\"ive thresholding 
(baseline) and the improved version using \ac{NMS}
(improved).
Groundtruth heatmaps of \textbf{\ac{hmap}} are generated using a Gaussian with 
$\sigma^2=0.1$.
This results in $\sim1$m of location uncertainty.
For \ac{hmap}, we use an output representation with a granularity of $10$cm.
To cope with the situation where the source is at the edge, the margin is added 
and results in a grid of $N=81$.
For simplicity, we choose \ac{tgrid} to have the same grid size $N$.
For \textbf{\ac{rgrid}}, the grid number $N=6$ and $(\lambda_1, 
\lambda_2) = (0.25, 10)$ in~\cref{eq:rg-loss}.

%
%
Second, we verify the effect of changes on the network architecture while fixing the output representation.
Three configurations of the network are compared.
\textbf{Single-encoder} is a baseline architecture where features of all microphones of all arrays are inputted to the encoder at once.
\textbf{Array-encoder} represents the architecture that separates the encoder for each array and shares their weights.
\textbf{Combined} is an architecture that uses both modifications by separating 
the encoder and rearranging channels by pairs of microphones. At the 
array-level, input features are duplicated and rearranged into a shape of 
$24$x$9$x$256$; [($R_1, R_2$)$, $($R_1, R_3$)$, ..., $($R_3, R_4$)$, $($I_1, I_2$)$, $($I_1, I_3$)$, ..., 
$($I_3, I_4$)].

%
%
\section{Experimental Results}
\label{ssec:results}

\begin{table*}[bt]
\footnotesize\addtolength{\tabcolsep}{-5pt}
\begin{center}
\caption{Output representation and architecture results}
\label{tab:resreparc}
\begin{tabular}{|c|c|cccccccccccccccccccccccc}
\hline
\multicolumn{2}{|c|}{Dataset}                                              & \multicolumn{8}{c|}{test-S0}                                                                                                                                                                                                                                                                                                   & \multicolumn{8}{c|}{test-A0}                                                                                                                                                                                                                                                                                                   & \multicolumn{8}{c|}{test-R0}                                                                                                                                                                                                                                                                                                   \\ \hline
\multicolumn{2}{|c|}{Resolution}                                           & \multicolumn{4}{c|}{0.3 m}                                                                                                                                & \multicolumn{4}{c|}{1 m}                                                                                                                                           & \multicolumn{4}{c|}{0.3 m}                                                                                                                                & \multicolumn{4}{c|}{1 m}                                                                                                                                           & \multicolumn{4}{c|}{0.3 m}                                                                                                                                & \multicolumn{4}{c|}{1 m}                                                                                                                                           \\ \hhline{|==========================|}
Output rep.              & DNN arch.                                              & \multicolumn{1}{c|}{Pre} & \multicolumn{1}{c|}{Rec} & \multicolumn{1}{c|}{\cellcolor[HTML]{EFEFEF}F1$\uparrow$} & \multicolumn{1}{c|}{\cellcolor[HTML]{EFEFEF}{\tiny RMSE}$\downarrow$} & \multicolumn{1}{c|}{Pre} & \multicolumn{1}{c|}{Rec} & \multicolumn{1}{c|}{\cellcolor[HTML]{EFEFEF}F1$\uparrow$} & \multicolumn{1}{c|}{\cellcolor[HTML]{EFEFEF}{\tiny RMSE}$\downarrow$}          & \multicolumn{1}{c|}{Pre} & \multicolumn{1}{c|}{Rec} & \multicolumn{1}{c|}{\cellcolor[HTML]{EFEFEF}F1$\uparrow$} & \multicolumn{1}{c|}{\cellcolor[HTML]{EFEFEF}{\tiny RMSE}$\downarrow$} & \multicolumn{1}{c|}{Pre} & \multicolumn{1}{c|}{Rec} & \multicolumn{1}{c|}{\cellcolor[HTML]{EFEFEF}F1$\uparrow$} & \multicolumn{1}{c|}{\cellcolor[HTML]{EFEFEF}{\tiny RMSE}$\downarrow$}          & \multicolumn{1}{c|}{Pre} & \multicolumn{1}{c|}{Rec} & \multicolumn{1}{c|}{\cellcolor[HTML]{EFEFEF}F1$\uparrow$} & \multicolumn{1}{c|}{\cellcolor[HTML]{EFEFEF}{\tiny RMSE}$\downarrow$} & \multicolumn{1}{c|}{Pre} & \multicolumn{1}{c|}{Rec} & \multicolumn{1}{c|}{\cellcolor[HTML]{EFEFEF}F1$\uparrow$} & \multicolumn{1}{c|}{\cellcolor[HTML]{EFEFEF}{\tiny RMSE}$\downarrow$}          \\ \hline
TG-rep (baseline)                &                            & .38                     & .87                     & \cellcolor[HTML]{EFEFEF}.53                    & \cellcolor[HTML]{EFEFEF}.15                      & .40                     & .92                     & \cellcolor[HTML]{EFEFEF}.56                    & \multicolumn{1}{c|}{\cellcolor[HTML]{EFEFEF}.23}          & .29                     & .34                     & \cellcolor[HTML]{EFEFEF}.31                    & \cellcolor[HTML]{EFEFEF}.20                      & .55                     & .65                     & \cellcolor[HTML]{EFEFEF}.59                    & \multicolumn{1}{c|}{\cellcolor[HTML]{EFEFEF}.44}          & .12                     & .24                     & \cellcolor[HTML]{EFEFEF}.16                    & \cellcolor[HTML]{EFEFEF}.19                      & .29                     & .56                     & \cellcolor[HTML]{EFEFEF}.38                    & \multicolumn{1}{c|}{\cellcolor[HTML]{EFEFEF}.50}          \\ \cline{1-1}
TG-rep (improved)                &                            & .86                     & .85                     & \cellcolor[HTML]{EFEFEF}.85                    & \cellcolor[HTML]{EFEFEF}.11                      & .95                     & \textbf{.94}            & \cellcolor[HTML]{EFEFEF}.94                    & \multicolumn{1}{c|}{\cellcolor[HTML]{EFEFEF}.20}          & .35                     & .35                     & \cellcolor[HTML]{EFEFEF}.35                    & \cellcolor[HTML]{EFEFEF}.20                      & .63                     & .64                     & \cellcolor[HTML]{EFEFEF}.64                    & \multicolumn{1}{c|}{\cellcolor[HTML]{EFEFEF}.45}          & .15                     & .25                     & \cellcolor[HTML]{EFEFEF}.19                    & \cellcolor[HTML]{EFEFEF}.19                      & .33                     & .57                     & \cellcolor[HTML]{EFEFEF}.42                    & \multicolumn{1}{c|}{\cellcolor[HTML]{EFEFEF}.47}          \\ \cline{1-1}
HM-rep                           &                            & \textbf{.94}            & \textbf{.88}            & \cellcolor[HTML]{EFEFEF}\textbf{.90}           & \cellcolor[HTML]{EFEFEF}\textbf{.10}             & \textbf{.99}            & .93                     & \cellcolor[HTML]{EFEFEF}\textbf{.96}           & \multicolumn{1}{c|}{\cellcolor[HTML]{EFEFEF}\textbf{.15}} & \textbf{.81}            & \textbf{.58}            & \cellcolor[HTML]{EFEFEF}\textbf{.68}           & \cellcolor[HTML]{EFEFEF}\textbf{.13}             & \textbf{.96}            & .69                     & \cellcolor[HTML]{EFEFEF}.80                    & \multicolumn{1}{c|}{\cellcolor[HTML]{EFEFEF}\textbf{.24}} & \textbf{.64}            & .39                     & \cellcolor[HTML]{EFEFEF}\textbf{.48}           & \cellcolor[HTML]{EFEFEF}\textbf{.15}             & \textbf{.89}            & .55                     & \cellcolor[HTML]{EFEFEF}\textbf{.67}           & \multicolumn{1}{c|}{\cellcolor[HTML]{EFEFEF}\textbf{.33}} \\ \cline{1-1}
RG-rep                           & \multirow{-4}{*}{Combined} & .91                     & .87                     & \cellcolor[HTML]{EFEFEF}.89                    & \cellcolor[HTML]{EFEFEF}\textbf{.10}             & .98                     & \textbf{.94}            & \cellcolor[HTML]{EFEFEF}\textbf{.96}           & \multicolumn{1}{c|}{\cellcolor[HTML]{EFEFEF}.17}          & .62                     & .55                     & \cellcolor[HTML]{EFEFEF}.58                    & \cellcolor[HTML]{EFEFEF}.17                      & .90                     & \textbf{.80}            & \cellcolor[HTML]{EFEFEF}\textbf{.85}           & \multicolumn{1}{c|}{\cellcolor[HTML]{EFEFEF}.33}          & .36                     & \textbf{.40}            & \cellcolor[HTML]{EFEFEF}.38                    & \cellcolor[HTML]{EFEFEF}.17                      & .60                     & \textbf{.66}            & \cellcolor[HTML]{EFEFEF}.63                    & \multicolumn{1}{c|}{\cellcolor[HTML]{EFEFEF}.40}          \\ \hline
                                 & Single-encoder             & .91                     & .85                     & \cellcolor[HTML]{EFEFEF}.88                    & \cellcolor[HTML]{EFEFEF}.10                      & \textbf{.99}            & .92                     & \cellcolor[HTML]{EFEFEF}.95                    & \multicolumn{1}{c|}{\cellcolor[HTML]{EFEFEF}.17}          & .77                     & .51                     & \cellcolor[HTML]{EFEFEF}.61                    & \cellcolor[HTML]{EFEFEF}.14                      & \textbf{.97}            & .64                     & \cellcolor[HTML]{EFEFEF}.77                    & \multicolumn{1}{c|}{\cellcolor[HTML]{EFEFEF}.26}          & .61                     & .37                     & \cellcolor[HTML]{EFEFEF}.46                    & \cellcolor[HTML]{EFEFEF}.16                      & .87                     & .52                     & \cellcolor[HTML]{EFEFEF}.65                    & \multicolumn{1}{c|}{\cellcolor[HTML]{EFEFEF}.37}          \\ \cline{2-2}
                                 & Array-encoder              & .93                     & .86                     & \cellcolor[HTML]{EFEFEF}.89                    & \cellcolor[HTML]{EFEFEF}\textbf{.09}             & \textbf{.99}            & .91                     & \cellcolor[HTML]{EFEFEF}.95                    & \multicolumn{1}{c|}{\cellcolor[HTML]{EFEFEF}\textbf{.15}} & .78                     & .57                     & \cellcolor[HTML]{EFEFEF}.66                    & \cellcolor[HTML]{EFEFEF}.15                      & \textbf{.97}            & \textbf{.71}            & \cellcolor[HTML]{EFEFEF}\textbf{.82}           & \multicolumn{1}{c|}{\cellcolor[HTML]{EFEFEF}.26}          & .61                     & \textbf{.41}            & \cellcolor[HTML]{EFEFEF}\textbf{.49}           & \cellcolor[HTML]{EFEFEF}\textbf{.15}             & .85                     & \textbf{.57}            & \cellcolor[HTML]{EFEFEF}\textbf{.68}           & \multicolumn{1}{c|}{\cellcolor[HTML]{EFEFEF}.35}          \\ \cline{2-2}
\multirow{-3}{*}{HM-rep}         & Combined                   & \textbf{.94}            & \textbf{.88}            & \cellcolor[HTML]{EFEFEF}\textbf{.90}           & \cellcolor[HTML]{EFEFEF}.10                      & \textbf{.99}            & \textbf{.93}            & \cellcolor[HTML]{EFEFEF}\textbf{.96}           & \multicolumn{1}{c|}{\cellcolor[HTML]{EFEFEF}\textbf{.15}} & \textbf{.81}            & \textbf{.58}            & \cellcolor[HTML]{EFEFEF}\textbf{.68}           & \cellcolor[HTML]{EFEFEF}\textbf{.13}             & .96                     & .69                     & \cellcolor[HTML]{EFEFEF}.80                    & \multicolumn{1}{c|}{\cellcolor[HTML]{EFEFEF}\textbf{.24}} & \textbf{.64}            & .39                     & \cellcolor[HTML]{EFEFEF}.48                    & \cellcolor[HTML]{EFEFEF}\textbf{.15}             & \textbf{.89}            & .55                     & \cellcolor[HTML]{EFEFEF}.67                    & \multicolumn{1}{c|}{\cellcolor[HTML]{EFEFEF}\textbf{.33}} \\ \hline
                                 & Single-encoder             & .82                     & .60                     & \cellcolor[HTML]{EFEFEF}.69                    & \cellcolor[HTML]{EFEFEF}.14                      & .97                     & .71                     & \cellcolor[HTML]{EFEFEF}.82                    & \multicolumn{1}{c|}{\cellcolor[HTML]{EFEFEF}.24}          & .67                     & .37                     & \cellcolor[HTML]{EFEFEF}.48                    & \cellcolor[HTML]{EFEFEF}.16                      & .92                     & .51                     & \cellcolor[HTML]{EFEFEF}.65                    & \multicolumn{1}{c|}{\cellcolor[HTML]{EFEFEF}.32}          & .54                     & .27                     & \cellcolor[HTML]{EFEFEF}.36                    & \cellcolor[HTML]{EFEFEF}.17                      & .86                     & .42                     & \cellcolor[HTML]{EFEFEF}.57                    & \multicolumn{1}{c|}{\cellcolor[HTML]{EFEFEF}.39}          \\ \cline{2-2}
                                 & Array-encoder              & .89                     & .68                     & \cellcolor[HTML]{EFEFEF}.77                    & \cellcolor[HTML]{EFEFEF}.12                      & \textbf{.98}            & .74                     & \cellcolor[HTML]{EFEFEF}.85                    & \multicolumn{1}{c|}{\cellcolor[HTML]{EFEFEF}\textbf{.19}} & \textbf{.75}            & .51                     & \cellcolor[HTML]{EFEFEF}.61                    & \cellcolor[HTML]{EFEFEF}.16                      & \textbf{.96}            & .66                     & \cellcolor[HTML]{EFEFEF}.78                    & \multicolumn{1}{c|}{\cellcolor[HTML]{EFEFEF}.28}          & \textbf{.60}            & .36                     & \cellcolor[HTML]{EFEFEF}\textbf{.45}           & \cellcolor[HTML]{EFEFEF}\textbf{.15}             & \textbf{.89}            & .53                     & \cellcolor[HTML]{EFEFEF}\textbf{.67}           & \multicolumn{1}{c|}{\cellcolor[HTML]{EFEFEF}\textbf{.36}} \\ \cline{2-2}
\multirow{-3}{*}{HM-rep (10\%*)} & Combined                   & \textbf{.89}            & \textbf{.71}            & \cellcolor[HTML]{EFEFEF}\textbf{.79}           & \cellcolor[HTML]{EFEFEF}\textbf{.11}             & \textbf{.98}            & \textbf{.78}            & \cellcolor[HTML]{EFEFEF}\textbf{.87}           & \multicolumn{1}{c|}{\cellcolor[HTML]{EFEFEF}\textbf{.19}} & .73                     & \textbf{.55}            & \cellcolor[HTML]{EFEFEF}\textbf{.63}           & \cellcolor[HTML]{EFEFEF}\textbf{.15}             & .94                     & \textbf{.71}            & \cellcolor[HTML]{EFEFEF}\textbf{.80}           & \multicolumn{1}{c|}{\cellcolor[HTML]{EFEFEF}\textbf{.27}} & .56                     & \textbf{.38}            & \cellcolor[HTML]{EFEFEF}\textbf{.45}           & \cellcolor[HTML]{EFEFEF}.16                      & .82                     & \textbf{.56}            & \cellcolor[HTML]{EFEFEF}\textbf{.67}           & \multicolumn{1}{c|}{\cellcolor[HTML]{EFEFEF}\textbf{.36}} \\ \hline
\end{tabular}
\begin{scriptsize}
\newline
\newline
\hspace*{0.3cm}{*model trained using 10\% of training set}\hspace*{\fill}Pre 
$\rightarrow$ Precision; Rec $\rightarrow$ Recall\hspace*{0.6cm}
\end{scriptsize}
\end{center}
\vspace{-1.0em}
\end{table*}

The output representation and architecture design results are given 
in~\cref{tab:resreparc}. Evaluation metrics were computed for resolutions of 
$0.3$ and $1.0$ meter for synthetic, augmented real and real test sets.

\textbf{Output Representation Comparison}
With a na\"ive keypoint retrieval approach (only thresholding), \ac{tgrid} 
(baseline) showed a competitive recall but with a poor precision across all 
test sets.
The weighted loss introduced in~\cref{eq:tg-loss} encourages the model to make multiple guesses at the vicinity of each expected source location.
Therefore, the number of \ac{FP} can be high.
The proposed keypoint retrieval method greatly enhances the localization performance in \ac{tgrid} (improved).
Nonetheless, the two introduced output representations (\ac{rgrid} and 
\ac{hmap}) consistently outperform the \ac{tgrid} output representation by a 
large margin.
Both introduced output representations have similar performance.
However, out of the two, \ac{hmap} seems to better generalize across the various testing settings.
For this reason, \ac{hmap} has been chosen for the comparison of architecture 
designs.

\textbf{Architecture Design Comparison}
This comparison is done on two training settings: full training set and limited 
training set (10\%), 
while all test sets are identical.
The results show that our proposed changes to the architecture improve the 
overall localization performance. The performance gap is relatively small with 
the full training set; however, with the limited training set, the margin 
between single-encoder and array-encoder architectures is significant.
This suggests that the proposed architecture requires less data to train.
The combined architecture does not always outperform the array-encoder 
architecture.
Both methods seem to lead to equivalent overall results most of the time.

\textbf{Impact of Resolution}
Looking at the output representation and architecture design results, we can 
see a rise in both precision and recall from fine resolution ($0.3$m) to 
coarser resolution ($1$m).
The reason is that more associations between ground truth and predicted 
keypoints can be made.
As unassociated predicted keypoints result in bad precision (\ac{FP} is high) 
and unassociated ground truth keypoints result in bad recall (\ac{FN} is high),
then augmenting the resolution leads to higher precision, recall and F1-score.
However, \ac{RMSE} increases with the resolution because keypoints can be further away. 

\begin{table}
\footnotesize
\addtolength{\tabcolsep}{-5pt}
\caption{Generalization towards the number of sources and musical genres (model trained with Combined architecture and HM-rep; metric calculated with $1$ meter resolution)} 
\label{tab:addlocper}
\subcaptionbox{Number of Sources}[.48\linewidth]{

\centering

\begin{tabular}{|c|c|cccc}
\hline
Dataset                   &  \# of Srcs      & \multicolumn{1}{c|}{Pre} & \multicolumn{1}{c|}{Rec} & \multicolumn{1}{c|}{\cellcolor[HTML]{EFEFEF}F1$\uparrow$} & \multicolumn{1}{c|}{\cellcolor[HTML]{EFEFEF}{\tiny RMSE}$\downarrow$} \\ \hline
                          & 1       & .99                     & .99                     & \cellcolor[HTML]{EFEFEF}.99                    & \multicolumn{1}{c|}{\cellcolor[HTML]{EFEFEF}.08} \\ \cline{2-2}
\multirow{-2}{*}{test-S0} & 2       & .98                     & .89                     & \cellcolor[HTML]{EFEFEF}.93                    & \multicolumn{1}{c|}{\cellcolor[HTML]{EFEFEF}.18} \\ \cline{1-2}
test-S1                   & 3       & .96                     & .64                     & \cellcolor[HTML]{EFEFEF}.77                    & \multicolumn{1}{c|}{\cellcolor[HTML]{EFEFEF}.22} \\ \hline
                          & 1       & .96                     & .81                     & \cellcolor[HTML]{EFEFEF}.88                    & \multicolumn{1}{c|}{\cellcolor[HTML]{EFEFEF}.22} \\ \cline{2-2}
\multirow{-2}{*}{test-A0} & 2       & .96                     & .63                     & \cellcolor[HTML]{EFEFEF}.76                    & \multicolumn{1}{c|}{\cellcolor[HTML]{EFEFEF}.25} \\ \cline{1-2}
test-A1                   & 3       & .98                     & .46                     & \cellcolor[HTML]{EFEFEF}.62                    & \multicolumn{1}{c|}{\cellcolor[HTML]{EFEFEF}.27} \\ \hline
                          & 1       & .90                     & .81                     & \cellcolor[HTML]{EFEFEF}.85                    & \multicolumn{1}{c|}{\cellcolor[HTML]{EFEFEF}.26} \\ \cline{2-2}
\multirow{-2}{*}{test-R0} & 2       & .87                     & .39                     & \cellcolor[HTML]{EFEFEF}.54                    & \multicolumn{1}{c|}{\cellcolor[HTML]{EFEFEF}.40} \\ \cline{1-2}
test-R1                   & 3       & .87                     & .32                     & \cellcolor[HTML]{EFEFEF}.46                    & \multicolumn{1}{c|}{\cellcolor[HTML]{EFEFEF}.42} \\ \hline\end{tabular}
}
\quad
\subcaptionbox{Musical Genres}[.48\linewidth]{

\centering
\begin{tabular}{|l|cccc}
\hline
Dataset (genre) & \multicolumn{1}{c|}{Pre} & \multicolumn{1}{c|}{Rec} & \multicolumn{1}{c|}{\cellcolor[HTML]{EFEFEF}F1$\uparrow$} & \multicolumn{1}{c|}{\cellcolor[HTML]{EFEFEF}{\tiny RMSE}$\downarrow$} \\ \hline
test-S0 (c-f)  & .99                     & .93                     & \cellcolor[HTML]{EFEFEF}.96                    & \multicolumn{1}{c|}{\cellcolor[HTML]{EFEFEF}.15} \\ \cline{1-1}
test-S2 (jazz)               & .99                     & .95                     & \cellcolor[HTML]{EFEFEF}.97                    & \multicolumn{1}{c|}{\cellcolor[HTML]{EFEFEF}.13} \\ \hline
test-A0 (c-f)   & .96                     & .69                     & \cellcolor[HTML]{EFEFEF}.80                    & \multicolumn{1}{c|}{\cellcolor[HTML]{EFEFEF}.24} \\ \cline{1-1}
test-A2 (jazz)               & .92                     & .54                     & \cellcolor[HTML]{EFEFEF}.68                    & \multicolumn{1}{c|}{\cellcolor[HTML]{EFEFEF}.37} \\ \hline
test-R0 (c-f)  & .89                     & .55                     & \cellcolor[HTML]{EFEFEF}.67                    & \multicolumn{1}{c|}{\cellcolor[HTML]{EFEFEF}.33} \\ \cline{1-1}
test-R2 (jazz)               & .92                     & .54                     & \cellcolor[HTML]{EFEFEF}.68                    & \multicolumn{1}{c|}{\cellcolor[HTML]{EFEFEF}.39} \\ \hline
\end{tabular}
\begin{scriptsize}
\newline
\newline
\hspace*{\fill}Pre $\rightarrow$ Precision; Rec $\rightarrow$ Recall
\end{scriptsize}

}
\vspace{-1.4em}
\end{table}

\textbf{Generalization on Source Number and Musical Genres} Additional 
localization performance results for one, two, and three sources as well as 
musical genres can be found in~\cref{tab:addlocper}.
Results are given for the best output representation (\ac{hmap}), best 
performing model (Combined) and resolution of $1$ meter because it seems more 
suitable as pairs of keypoints tend to be around $0.3$ meters away on average 
for tests on real data with a resolution of $1$ meter.
Even though models were trained with 1 and 2 sources, the results for 3 
simultaneous 
sources are still relatively good compared to the cases where there are 1 and 2 
sources.
A model trained on synthetic dataset (train-S) has similar performance on 
\textit{classical-funk} and \textit{jazz} musical genres.
But a model trained on real world with augmentation dataset (train-A) does not 
generalize as much to other musical genres.

\textbf{Synthetic, Augmented and Real World Data Comparison}
Despite the significant recording and labeling effort, the models trained on 
real world data still lack diversity compared to the synthetic environment.
This is also reflected by the performance drop from synthetic to real world with augmentation.
An additional performance decrease is observed from the real world with 
augmentation to the real world on \textit{classical-funk} data.
However it is not the case with \textit{jazz} data.
The reason is that \textit{classical-funk} data contains lots of variations in loudness and silent pauses while \textit{jazz} tends not to.
In the real world with augmentation data, those weak intensity passages (where 
sources cannot be considered active) are filtered out whereas with real world 
data this is not possible.

%
%
\section{Conclusion}
This paper introduced a novel architecture as well as two novel 
output representations for deep learning based multiple sound source 2D 
localization. These lead to a significant improvement of the localization 
performance as shown by the extensive experiments conducted on both simulated 
and real-world data.
A remaining issue is the performance degradation from the simulated 
environment to the real world.
This is probably due to the insufficient amount of real 
world data for training a DNN.
In the future, we plan to leverage simulation to 
inexpensively generate a large amount of labeled data and then train the 
localization model in such a way that the knowledge is transferable to the real 
world.

%
%


\bibliographystyle{IEEEtran}
\bibliography{IEEEabrv,reference}

\begin{thebibliography}{10}
\providecommand{\url}[1]{#1}
\csname url@samestyle\endcsname
\providecommand{\newblock}{\relax}
\providecommand{\bibinfo}[2]{#2}
\providecommand{\BIBentrySTDinterwordspacing}{\spaceskip=0pt\relax}
\providecommand{\BIBentryALTinterwordstretchfactor}{4}
\providecommand{\BIBentryALTinterwordspacing}{\spaceskip=\fontdimen2\font plus
\BIBentryALTinterwordstretchfactor\fontdimen3\font minus
  \fontdimen4\font\relax}
\providecommand{\BIBforeignlanguage}[2]{{%
\expandafter\ifx\csname l@#1\endcsname\relax
\typeout{** WARNING: IEEEtran.bst: No hyphenation pattern has been}%
\typeout{** loaded for the language `#1'. Using the pattern for}%
\typeout{** the default language instead.}%
\else
\language=\csname l@#1\endcsname
\fi
#2}}
\providecommand{\BIBdecl}{\relax}
\BIBdecl

\bibitem{conf:nips-ws:chakrabarty2017}
S.~Chakrabarty and E.~A. Habets, ``Multi-speaker localization using
  convolutional neural network trained with noise,'' in \emph{Workshop Machine
  Learning for Audio Signal Processing at {NIPS} ({ML}4Audio@{NIPS}17)}.

\bibitem{conf:icra:he2018}
W.~{He}, P.~{Motlicek}, and J.~{Odobez}, ``Deep neural networks for multiple
  speaker detection and localization,'' in \emph{2018 IEEE International
  Conference on Robotics and Automation (ICRA)}, May 2018, pp. 74--79.

\bibitem{conf:interspeech:he2018}
W.~He, P.~Motlicek, and J.-M. Odobez, ``Joint localization and classification
  of multiple sound sources using a multi-task neural network,'' in \emph{Proc.
  Interspeech 2018}, 2018, pp. 312--316.

\bibitem{jour:wcmc:cobos2017}
M.~Cobos, F.~Antonacci, A.~Alexandridis, A.~Mouchtaris, and B.~Le, ``A survey
  of sound source localization methods in wireless acoustic sensor networks,''
  \emph{Wireless Communications and Mobile Computing}, 2017.

\bibitem{jour:signal:ma2006}
{Wing-Kin Ma}, {Ba-Ngu Vo}, S.~S. {Singh}, and A.~{Baddeley}, ``Tracking an
  unknown time-varying number of speakers using tdoa measurements: a random
  finite set approach,'' \emph{IEEE Transactions on Signal Processing},
  vol.~54, no.~9, pp. 3291--3304, Sep. 2006.

\bibitem{jour:tasl:Alexandridis2018}
A.~Alexandridis and A.~Mouchtaris, ``Multiple sound source location estimation
  in wireless acoustic sensor networks using doa estimates: The
  data-association problem,'' \emph{IEEE/ACM Trans. Audio, Speech and Lang.
  Proc.}, vol.~26, no.~2, pp. 342--356, Feb. 2018.

\bibitem{jour:sp:griffin2015}
A.~Griffin, A.~Alexandridis, D.~Pavlidi, Y.~Mastorakis, and A.~Mouchtaris,
  ``Localizing multiple audio sources in a wireless acoustic sensor network,''
  \emph{Signal Processing}, vol. 107, pp. 54 -- 67, 2015.

\bibitem{jour:aslp:sundar2018}
H.~{Sundar}, T.~V. {Sreenivas}, and C.~S. {Seelamantula}, ``Tdoa-based multiple
  acoustic source localization without association ambiguity,'' \emph{IEEE/ACM
  Transactions on Audio, Speech, and Language Processing}, vol.~26, no.~11, pp.
  1976--1990, Nov 2018.

\bibitem{tech:dcase:Inoue2018}
T.~Inoue, P.~Vinayavekhin, S.~Wang, D.~Wood, N.~Greco, and R.~Tachibana,
  ``Domestic activities classification based on {CNN} using shuffling and
  mixing data augmentation,'' DCASE2018 Challenge, Tech. Rep., September 2018.

\bibitem{conf:slt:takeda2016}
R.~{Takeda} and K.~{Komatani}, ``Discriminative multiple sound source
  localization based on deep neural networks using independent location
  model,'' in \emph{2016 IEEE Spoken Language Technology Workshop (SLT)}, Dec
  2016, pp. 603--609.

\bibitem{jour:sensors:diaz2018}
J.~Vera-Diaz, D.~Pizarro, and J.~Macias-Guarasa, ``Towards end-to-end acoustic
  localization using deep learning: From audio signals to source position
  coordinates,'' \emph{Sensors}, vol.~18, no.~10, p. 3418, Oct 2018.

\bibitem{conf:mlsp:vesperini2016}
F.~{Vesperini}, P.~{Vecchiotti}, E.~{Principi}, S.~{Squartini}, and
  F.~{Piazza}, ``A neural network based algorithm for speaker localization in a
  multi-room environment,'' in \emph{2016 IEEE 26th International Workshop on
  Machine Learning for Signal Processing (MLSP)}, Sep. 2016, pp. 1--6.

\bibitem{jour:stsp:adavanne2019}
S.~{Adavanne}, A.~{Politis}, J.~{Nikunen}, and T.~{Virtanen}, ``Sound event
  localization and detection of overlapping sources using convolutional
  recurrent neural networks,'' \emph{IEEE Journal of Selected Topics in Signal
  Processing}, vol.~13, no.~1, pp. 34--48, March 2019.

\bibitem{conf:eusipco:adavanne2018}
S.~{Adavanne}, A.~{Politis}, and T.~{Virtanen}, ``Direction of arrival
  estimation for multiple sound sources using convolutional recurrent neural
  network,'' in \emph{2018 26th European Signal Processing Conference
  (EUSIPCO)}, Sep. 2018, pp. 1462--1466.

\bibitem{conf:cvpr:redmon2016}
J.~{Redmon}, S.~{Divvala}, R.~{Girshick}, and A.~{Farhadi}, ``You only look
  once: Unified, real-time object detection,'' in \emph{2016 IEEE Conference on
  Computer Vision and Pattern Recognition (CVPR)}, June 2016.

\bibitem{conf:icassp:scheibler2018}
R.~Scheibler, E.~Bezzam, and I.~Dokmanic, ``Pyroomacoustics: A python package
  for audio room simulation and array processing algorithms,'' in \emph{2018
  IEEE International Conference on Acoustics, Speech, and Signal Processing,
  ICASSP 2018 - Proceedings}, 9 2018.

\bibitem{conf:mmsp:Alexandridis2018}
A.~{Alexandridis}, A.~{Griffin}, and A.~{Mouchtaris}, ``Multiple source
  location estimation on a dataset of real recordings in a wireless acoustic
  sensor network,'' in \emph{2018 IEEE 20th International Workshop on
  Multimedia Signal Processing (MMSP)}, Aug 2018, pp. 1--6.

\end{thebibliography}

\end{document}